# Flexible Spectro Interferometric modelling of OIFITS data with PMOIRED[*]

Antoine Mérand[†a]

[a]European Southern Observatory, Garching bei München, Germany

## ABSTRACT

Despite image reconstruction becoming more widespread when interpreting OIFITS Data, model fitting in u,v space often remains the best way to interpret data, either because of the sparsity of the data, or because a quantitative measurement needs to be done. PMOIRED, is a flexible Python library to visualize, manipulate and model OIFITS data using simple geometric models. The strength of PMOIRED resides in its capability to combine linearly various simple components to create complex scenes, while linking, constraining, and adding priors to fitted parameters. The code also enables grid search to find global minima, as well as data resampling to better evaluate uncertainties. In addition to analytical functions, arbitrary radial profiles, azimuthal variations or sparse wavelet modelling of spectra are implemented.

## 1. INTRODUCTION

Image reconstruction, even for a visibility dataset with a rich u,v coverage, is still a complex process, and it is often complicated to estimate the statistical properties of the reconstructed image (for example, the uncertainties on the images). Even at the age of u,v rich data sets, forward modeling and fitting of the visibility is still a must for many astrophysical observations, in particular when one wants to extract physical parameters. Instead of reconstructing an image, parametric modeling assumes the morphology of the object leads to an image which can be modeled with only a few parameters.

For instance, when observing a simple star (single, non-rotating, not surrounded by a disk nor with substantial mass loss etc.), it is reasonable to model the interferometric observations with a filled disk, with a radial luminosity variation to account for the center-to-limb darkening (CLD) which occurs in real stars, due to the temperature gradient in the photosphere. This realistic parametrization, when informed by stellar photospheric models, allows to derive physical quantities such as the Rosseland radius (optical depth of 2/3 in the photosphere) of the star or CLD coefficients, which can be compared to the model's predictions. Conversely, a departure from a CLD disk indicates that the star is not simple. Parametric modeling can still be used to model rotationally induced gravity darkening, the presence of disk, circum-stellar envelops etc.

Finally, one can argue that optical interferometers have become open, accessible and robust in terms of observations preparation, observations execution, data reduction and image reconstruction. Data modeling, especially when it comes to spectro-interferometry, is not at this level, hence the initiative to develop a new data analysis tool presented here: PMOIRED (Parametric Modeling of Optical InteRferomEtric Data) .

## 2. REQUIREMENTS

PMOIRED was developed with the following requirements. These requirements are almost all met by currently available tools such as LitPRO [1], except notably for spectro-interferometric modeling:
- It must be easy to use and robust for non-experts: the users' interactions should not require advanced knowledge

---

[*] http://ascl.net/code/v/3286
[†] amerand@eso.org; https://www.eso.org/~amerand

- Fit all data in the OIFITS format [2]: amplitude and squared amplitude of visibilities, closure phases, differential phases and normalised spectra (as in "in a spectral line"), as well as spectral energy distribution.
- Include all necessary data manipulation such as seamlessly combining several data files (including from different instruments and/or different observing modes), manage flagged data and manipulate error bars
- Advanced spectro-interferometric features: model in emission / absorption lines, computation of differential quantities and correction for tellurics (in spectra)
- Easy plotting, as well as access to underlying data for custom plots
- Manipulate data in "human times": execution should be measured in seconds or minutes, not hours or days. It is important to be interact with data and try easily new ideas. Use parallel programming when possible
- Science case agnostic: the syntax to describe the model should address generic morphologies rather than using astrophysical modelling (e.g. radiative transfer)

For these reasons, the code was developed in Python, which is the most used language in the astronomical community for data analysis. In order to be fast, modelling will rely on using simple analytical functions to describe morphologies such as disks, rings and Gaussians. The complexity of the model is achieved by combining linearly many components since the modelled quantity, the complex visibility, is a linear combination of the Fourier Transform of the components.

Using analytical visibilities is, in practice, thousands of times faster than using numerical Fourier Transform (FT) of synthetic images, even when using algorithms such as the Fast Fourier Transform. Moreover, using numerical Fourier Transform requires careful consideration to maintain accuracy which depends on the spatial frequencies probed.

The syntax to construct model is using Python dictionaries, which are data structure akin to lists, with elements indexed by any hashable variable. In the case of PMOIRED model, the keys to the dictionary is a string defining a parameter of the model. The value of the parameter can be a number, of a string, as we will describe later.

## 3.  MODELING BUIDLING BLOCKS

### 3.1  Components with analytical visibilities

The basic building blocks with analytical visibilities are the uniform disk, and gaussians, as well as the special trivial cases include the unresolved component (V=1) and the fully resolved component (V=0). Assuming we observe at baseline $B = \sqrt{u^2 + v^2}$ where (u,v) is the projection baseline vector, and wavelength $\lambda$:

- Uniform disk of angular diameter $\theta$: $V(B,\lambda) = 2\frac{J_1(x)}{x}$ where $x = \pi B\theta/\lambda$ and $J_1$ is the Bessel function. In PMOIRED, it will be modeled as `{'ud':1.0}` for a 1.0mas (milliarcseconds) uniform disk.

- Gaussian of full width at half maximum $\alpha$: $V(B,\lambda) = e^{-(\frac{2\ln(2)B}{\lambda a})^2}$. In PMOIRED, it will be modeled as `{'fwhm':2.0}` for a 2.0mas Gaussian.

Additional complexity can be added by applying geometrical transform. A displacement in the image plane of (x,y) corresponds to multiplying the complex visibility by $e^{-i\phi}$ where $\phi = 2\pi(xu + yv)/\lambda$; a rotation in the direct space (x,y) corresponds to similar rotation in the spatial frequency space (u,v). Finally, a stretch o factor "a" along one axis corresponds to a stretch of 1/a in the corresponding spatial frequency direction. By combining rotation and stretch along one axis, one can reproduce the inclination and projection angle of an inclined disk for instance.

### 3.2  Derived analytical visibilities

Two uniform disk, one positive and one (smaller) negative can be combined to form a ring. If the inner disk is offset, it will result in a skewed ring. `{'diamin':1.0, 'diamout':2.0}` will be a ring of 1.0mas of inner diameter and 2.0 outer diameter.

### 3.3  Semi-analytical visibilities

The components above are not sufficient to describe the complexity of astrophysical objects. For instance, one expects a circum-stellar disk to have an intensity profile which varies with radial distance to the central star. Conversely, a stellar disk is not a uniform disk but displayed central-to-limb darkening which is dictated, among other things, by the temperature gradient and optical depth of the photosphere. The exact profile depends on the astrophysical processes. To model this, PMOIRED allows to describe arbitrarily the profiles of disks and rings. The resulting visibility is the (1D) Hankel transform of the intensity profile I(r):

$$V = \left.\int I(r) J_0(2\pi Br/\lambda)\, r dr \middle/ \int I(r) r dr \right.$$

where $J_0$ is the 0-th order Bessel function. A power law central to limb darkening [3] will be modeled as `{'diam':1.0, 'profile':'$MU**0.2'}` where diam is the diameter in mas, and $MU represent the cosine of the emerging ray of the stellar surface: $\mu = \sqrt{1 - (r/r_{max})^2}$ where r is the apparent distance from a point of the apparent stellar disk to the center of the disk.

A ring with an exponential decreasing intensity will be modeled as `{'diamin':1.0, 'diamout':2.0, 'profile':'np.exp(-$R)'}` where $R is the radial distance (in mas).

Note that 1) 'profile' is a string which should be executable in Python and 2) the absolute value of the intensity profile does not encode the brightness of the component, as the 1D visibility transformed is normalized to the total visibility.

Centro-symmetric astrophysical objects not only display radial brightness variation, but also azimuthal. For instance, the inner rim of an inclined protoplanetary disk will display significant brightening on the exposed side, due to projection effects. If the brightness of an image can be described in polar projection as a product of functions, one strictly radial and one strictly azimuthal harmonic, i.e. $F(r,\theta) = I(r) \sum_{j \geq 0} a_j \cos(j[\theta + \phi_j])$, the Fourier transform can be decomposed as follow:

$$V(B, \psi) = 2\pi \sum_{j \geq 0} a_j (-i)^j H_j(B) \cos(j[\psi + \phi_j])$$

Where $H_j(B)$ is the j-th order Hankel transform of I(r) (i.e. same as above, but replacing the 0-th Bessel function by the j-th) and $\psi$ is the baseline's projection angle, i.e. the Fourier conjugate variable of the direct space projection angle $\theta$. This is a generalization of the case described in [4], where it is assumed that I(r) is a delta function.

### 3.4 Multi-components

Combining components is achieved in a single dictionary. For instance, a central star with a disk will be modeled as:

```
{'star,ud':0.1,          # uniform disk of 1mas diameter
 'disk,diamin':1.0,      # disk inner diameter (in mas)
 'disk,diamout':2.0,     # disk outer diameter (in mas)
 'disk,profile':'$R**-2',# disk radial profile
 'disk,x':-0.1,          # disk offset along u (in mas)
 'disk,y':0.3,           # disk offset along v (in mas)
 'disk,f':2.2,           # disk total flux (arb unit)
 'disk,incl": 50,        # inclination of the disk (in degrees, face-on=0)
 'disk,projang':45       # projection angle (in degrees, N=0, E=90)
}
```

5 new parameters are introduced: 'x', 'y' (image plane shift), 'f' (total flux), 'incl' (inclination) and 'projang' (projection angle). These parameters have default values x=0, y=0, f=1, incl=0.

The name of the components, 'star' and 'disk' are arbitrary and are not used by PMOIRED to determine the nature of the component, only the parameters used are. This also means that many of the same components can be used (e.g. several rings)

### 3.5 Parameters self-reference

Parameters can be self-referenced. For instance, for the previous example, one could express the inner diameter of the disk as a multiple of the stellar diameter. This is achieved by defining: `{'disk,diamin':'5*$star,diam'}`.

One can also introduce additional parameters. For instance, the power law of the disk's radial profile can be parametrized with a variable (instead of a value): `{'disk,profile':'$R**$powerlaw', 'powerlaw':-2}`. This allows to fit the power law index, or to have several components to share a parametrization (e.g. nested rings with all same inclination and projection angle). This powerful syntax allows to describe complex objects which visibility is still computed analytically (of semi-analytically) and using few free parameters. Figure 1 shows some examples of simple to complex models. A Python Notebook is provided‡ with the tool to show all the possible models.

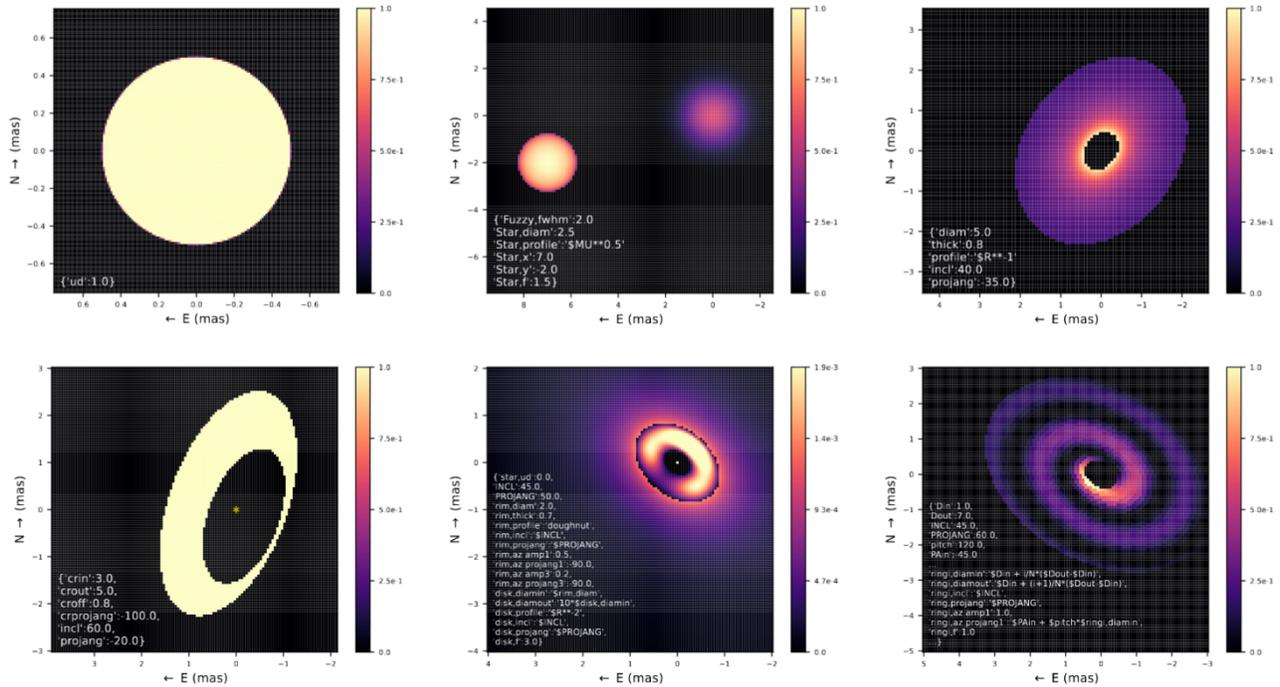

*Figure 1 Gallery of models using the PMOIRED syntax, from simple components to complex combination including self-reference to parameters.*

### 3.6 Spectral energy distribution and normalized spectra

Using multi-components require to define the relative fluxes between the components. The most basic definition is to use the parameter `'f'` for each component, which designates the total flux of each component. By default (i.e. if omitted), `'f':1` is used. To introduce a wavelength dependency, one should use for example `'spectrum':'$WL**-2'` where $WL is the reserved PMOIRED variable designating the wavelength (in microns). Note that contrary to `'profile'`, the absolute value of `'spectrum'` matters. So a star+disk model could be parametrized by not defining the stellar's flux, but having {`'disk,spectrum':'$F0*$WL**$FPOW'`, `'F0':0.5`, `'FPOW':2`}. Since the stellar flux is by default 1 for every wavelength, `'disk,spectrum'` is essentially the flux ratio between the disk and the star. Since the visibility is not sensitive to the absolute flux, a reference aways need to be set either in the definition of the model, or by fixing some parameters during the fit (see below).

Emission and absorption lines can be described using a defined syntax to parametrize Gaussian or Lorentzian profiles. PMOIRED is optimized in several ways to handle spectral lines. First, it is possible to fit the normalize/differential quantities, such as the normalized flux (i.e. the flux divided by the flux in the continuum) and the differential phase (i.e. the visibility's phase, to which the phase in the continuum has been subtracted). PMOIRED assumes the continuum by using the analytical definition of emission/absorption lines: the same definition is used to adjust the continuum to the data and the model, ensuring the fit is robust to the definition used to compute the differential phase in the OIFITS file. Second, it is important to correct the high resolution spectra (R>1000) from telluric features: the features do not affect the interferometric quantities derived from the visibility, but does affect the observed spectra. PMOIRED offers such correction for the GRAVITY data.

---

‡ https://github.com/amerand/PMOIRED/blob/master/examples/Model%20definitions%20and%20examples.ipynb

# 4. FITTING MODELS TO DATA

The simplest PMOIRED session to fit a model to data has the following steps:
- Load data file(s) with the constructor: `oi=pmoired.OI(…)`
- Set the context of the fit (e.g which observable to fit, how to scale error bars etc): `oi.setupFit(…)`
- The fit itself: `oi.doFit(…)`
- Displaying the result: `oi.show(…)`

The bare minimum requires to explicitly define which observable to fit, and the initial model (and parameter's value) for the fit. The exact syntax is given in the examples provided[§] with PMOIRED.

## 4.1 Setting the fit context

Setting the context of the fit is mandatory in PMOIRED: at least a list of the observable to be fitted needs to be provided. In addition, basic manipulation of error bars can be done. Error bars for interferometric observations are hard to estimate, and unrealistic low uncertainties can throw off a $\chi^2$ minimization into a local minimum. PMOIRED let the user define minimum uncertainties for observations (i.e. will change the error bars to this value), define maximum uncertainties (i.e. will exclude data with larger error bars), as well as define multiplication factors, which can be useful to increase/decrase the relative weight of certain data.

```
setupFit({'obs':['V2', 'T3PHI'],
        'min relative error':{'V2':0.01},
        'min error':{'T3PHI':1.5}})
```

means that only squared visibilities (V2) and closure phases (T3PHI) will be fitted, and a minimum of 1% (relative) error for V2 and 1.5 degrees in T3PHI is enforced.

The OIFITS format allows to flag bad data at the data reduction process. Even if feasible, PMOIRED does not provide (yet) easy tools to access the data and flag individual visibility data points.

## 4.2 Fitting parameters and results

The fit is based on a $\chi^2$ minimization, and hence require an initial guess from which it will use a gradient descent (`scipy.minimisation.leastsq`). The algorithm is however accessed using a custom high level library which allows to fix some parameters, as well as set priors for the fitted one. The result of the fit is not only the best fit parameters (and local minimum $\chi^2$ value), but also the uncertainties and correlation matrix, both derived from the covariance matrix. The correlation matrix is important to make sure parameters are independent from each other. A high correlation (e.g. superior to ~90%) indicates that parameters are degenerate, either because of a poor parametrization of the model or because the data cannot distinguish between the two parameters. There are no general rules to understand and fix correlated parameters.

Additional advanced tools are provided to determine the global best fit values, as well as evaluate realistic uncertainties:

- `bootstrapFit`: this performs several fits (typically dozens to hundreds) after having resampled the data. Since interferometric data are highly correlated [5], but these correlations seem not to bias the best fit, rather the uncertainties [6]. It is important to resample data by drawing/ignoring spectral data taken at a same date and with the same baseline, since the highest correlation is expected among such data.
- `gridSearch`: provides ways to explore the initial parameters space using a grid and/or randomization in a certain range.

Both methods are highly parallel problem: PMOIRED will use multi-cores CPU at their best to accelerate these 2 processes.

# 5. COMPLETE EXAMPLE

Let's take the case of GRAVITY spectro-interferometric observation of a binary [*typical computing times are indicated for each step*]:

---

[§] https://github.com/amerand/PMOIRED/tree/master/examples

1. Determining the separation of the binary: data are loaded and binned spectrally. A grid search is performed using an approach like in CANDID[**] [7]. The fitted data and the visibility amplitude and closure phases. [*20 seconds*]
2. The GRAVITY spectra are corrected to correct for the telluric absorption lines. The approach is similar to Molecfit[††] [8], an uses a grid of telluric models from the same tool (A. Smette, private communication). The result in written in the OIFITS files for later use: this operation is only required to be ran once [*100 seconds*]
3. The model for the full fit is initialized using the separation vector and contrast ratio of the binary found using the grid search, now including spectral lines for each of the components, at the 3 wavelengths where one sees lines in the spectrum of spectro-interferometric quantities. A global fit is performed on all parameters and all data: normalized spectrum, closure phases, visibility amplitude and differential phases. Note that the continuum for the normalized spectrum and differential phase is implicitly defined by the spectral lines of the model and is recomputed at each iteration of the minimization. 18 free parameters are fitted. [*7s in total, 23ms for each iteration*]
4. Once the least-square algorithm has found the best fit solution, data resampling (bootstrapping) is used to better estimate the uncertainties. [*180 seconds*]

The whole analysis requires only a few minutes. Each step is illustrated on Figure 2, the result of the fit is illustrated on Figure 3.

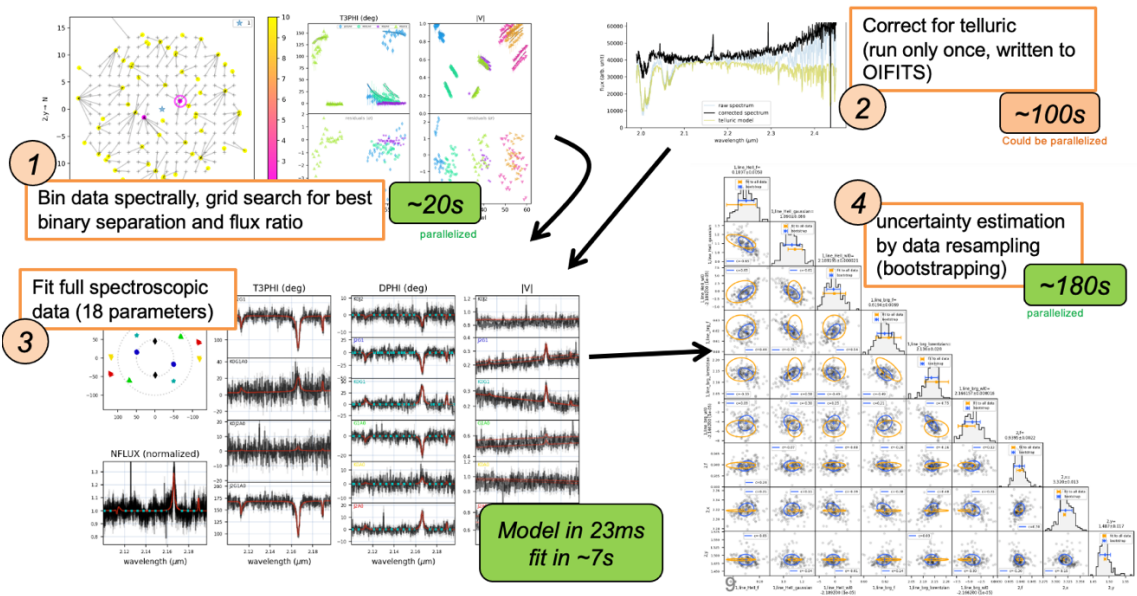

*Figure 2 Illustration of a spectro-interferometric fit of the GRAVITY observations of a binary star, displaying emission lines. 1) grid fit on binned data to find the best binary separation and contrast ratio in the continuum 2) telluric correction of spectra 3) global spectro-interferometric fit including spectral lines 4) final estimation of parameters and uncertainties using data resampling (bootstrapping)*

---

[**] https://github.com/amerand/CANDID
[††] http://www.eso.org/sci/software/pipelines/skytools/molecfit

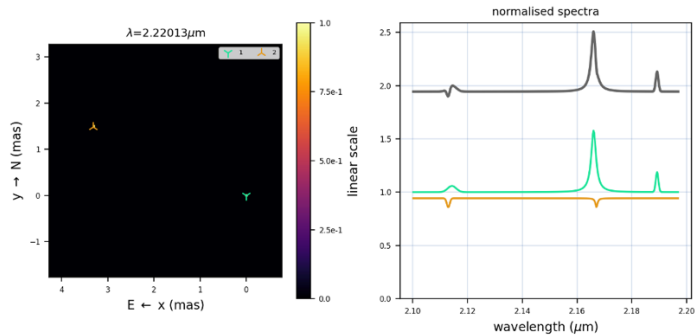

*Figure 3 Model of the binary star: left is a synthetic image showing the 2 unresolved stars in the field; right shows the normalized spectra of each star, as determined by the fit to the GRAVITY spectro-interferometric data (note that although stars have very similar fluxes in the continuum, only one star has emission lines whereas the other not: this is the important astrophysical result of the analysis).*

## 6. CONCLUSION AND PROSPECT

PMOIRED [9] is available is a complete solution to display and model OIFITS data, using parametric modeling. The observed object can be modeled spatially and spectrally using a combination of simple elements, taking advantage of the linearity of the Fourier Transform and the know analytical and semi-analytical formulas of simple objects such as disks, rings and Gaussians etc. PMOIRED has enabled 5 refereed articles in the past 12 months, showing the need for an easily accessible and robust analysis tool for (spectro-) interferometric data. PMOIRED is being continuously developed to include new features such as time-varying parameters (e.g. to fit orbital motion), case specific models (e.g. Keplerian disk, microlensing).

Another important line of development is data simulation, based on astrophysical data cubes (images as function of wavelength). This feature has several applications: it can be used to simulate data in order to assess the feasibility of an observing program, or the detectability of a certain phenomenon. Another application of data synthesis is parametric retrieval of astrophysical quantities. Parametric morphological modeling uses parameters that may be related to an astrophysical quantity but is not accurate. A basic example of this concept is the angular diameter of stars: even if a uniform disk diameter can be determined very precisely with a model barely departing from the interferometric data, the uniform disk diameter needs to be corrected for the central-to-limb darkening to determine the Rosseland diameter. Using PMOIRED, the course of action would be to use stellar atmosphere models to generate realistic images of stars (i.e. limb-darkened) with a known angular Rosseland diameter, synthesis the interferometric observations and fit a analytical model. One can study the accuracy of the angular diameter estimation compared to the input one.